\documentclass{PoS}
\usepackage{lineno}
\title{The MAPS-based ITS Upgrade for ALICE}
\ShortTitle{ALICE ITS2}

\author{\speaker{Giacomo Contin}\thanks{on behalf of the ALICE Collaboration.}\\
        Dipartimento di Fisica dell’Università and Sezione INFN, Trieste, Italy\\
        E-mail: \email{giacomo.contin@ts.infn.it}}
\author{on behalf of the ALICE Collaboration}

\abstract{The Inner Tracking System (ITS) Upgrade for the ALICE experiment
at LHC is the first large-area ($\sim$10~m$^2$) silicon vertex detector based on
the CMOS Monolithic Active Pixel Sensor (MAPS) technology, which
combines sensitive volume and front-end readout logic in the same piece
of silicon. This technology allows a reduced material budget (target value
of 0.3\% on the innermost layers) thanks to the thin sensors (50-100~$\mu$m)
and limited need of cooling, in combination with light-material interconnection circuits and support structures. The small pixel pitch ($\sim$30~$\mu$m),
the location of the layers (7 cylindrical layers with radii ranging from
2.3~cm to 39.3~cm from the beam interaction line), and the limited material budget will provide the ALICE experiment with extremely precise
tracking resolution. The high-rate readout capabilities will also enable
ALICE to collect a large data sample at the 50~kHz Pb--Pb collision rate
expected in the LHC Run~3.
The new ITS, now assembled at the surface, is currently undergoing
an exhaustive pre-commissioning phase with standalone calibration and
cosmic ray data-taking, which will be completed by April 2020 before the
installation in the ALICE detector. Experience gained from the construction and the pre-commissioning phase, and plans for the installation and
preparation for the data-taking in ALICE will be presented in this paper.
The role played by the new ITS within the development path of the MAPS
technology for future applications will also be briefly discussed.
}

\FullConference{The 28th International Workshop on Vertex Detectors - Vertex2019\\
		13-18 October, 2019\\
		Lopud, Croatia}

\begin{document}

\section{Introduction}
The Inner Tracking System (ITS) for ALICE (\emph{A Large Ion Collider Experiment}) at the Large Hadron Collider (LHC) will be upgraded during the second LHC Long Shutdown (LS2), in 
2019\slash2020. The ITS Upgrade ("ITS2") will be the first large-area silicon tracker based on the CMOS Monolithic Active Pixel Sensor (MAPS) technology operating at a collider.

The ITS2 layout \cite{ITS_TDR} includes 7 cylindrical layers grouped into two sub-systems: the Inner Barrel (IB), consisting of three 27~cm-long layers with radii of 2.3, 3.1 and 3.9~cm, respectively, and the Outer Barrel (OB), composed of two 84~cm-long Middle Layers (ML) placed at 24 and 30~cm from the interaction line, and two 148~cm-long Outer Layers (OL) at 42 and 48~cm, respectively. The layers cover a total surface area of $\sim$10~m$^2$ with 12.5~G pixels. 

\begin{figure}[htbp]
\centering
\includegraphics[width=.7\textwidth]{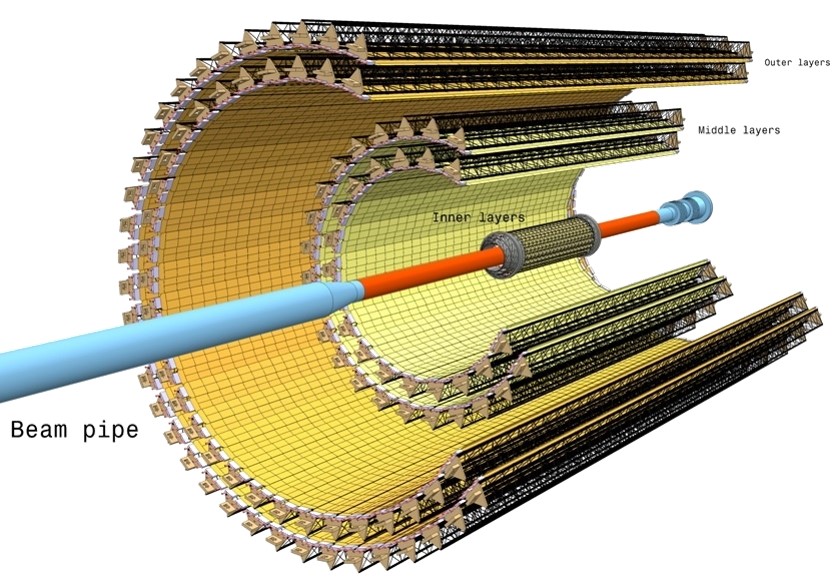}
\caption{\label{fig:PXL}The ITS2 layout, with three Inner layers close to the beam pipe, two Middle layers and two Outer layers.}
\end{figure}

The ITS2 is based on the ALPIDE chip, developed within the ALICE Collaboration for this application. It is produced in the TowerJazz 180~nm CMOS imaging process \cite{Markus}. It offers a deep PWELL, which can be used to shield the NWELL of PMOS transistors. This allows the use of full CMOS circuitry in the pixel area without the drawback of parasitic charge collection by the NWELLs. The process allows the use of a high-resistive ($>1$~k$\Omega\cdot$m) epitaxial layer on a p-substrate, which increases the radiation tolerance. In addition, applying a moderate negative voltage to the substrate can be used to increase the depletion zone around the collection diode and improve the signal-to-noise ratio.
The main characteristics of the sensor are:
\begin{itemize}
  \item pixel size: $\sim 29\times 27~\mu$m$^2$,
  \item thickness: 50~$\mu$m (inner layers) - 100~$\mu$m (outer layers),
  \item integration time: $<20~\mu$s,
  \item in-pixel discriminators and in-matrix address encoder with asynchronous sparsified readout,
  \item maximum readout speed: 1.2~Gbit/s, 
  \item power consumption: 40~mW/cm$^2$.
\end{itemize}

\begin{figure}[htbp]
\centering
\includegraphics[width=.435\textwidth]{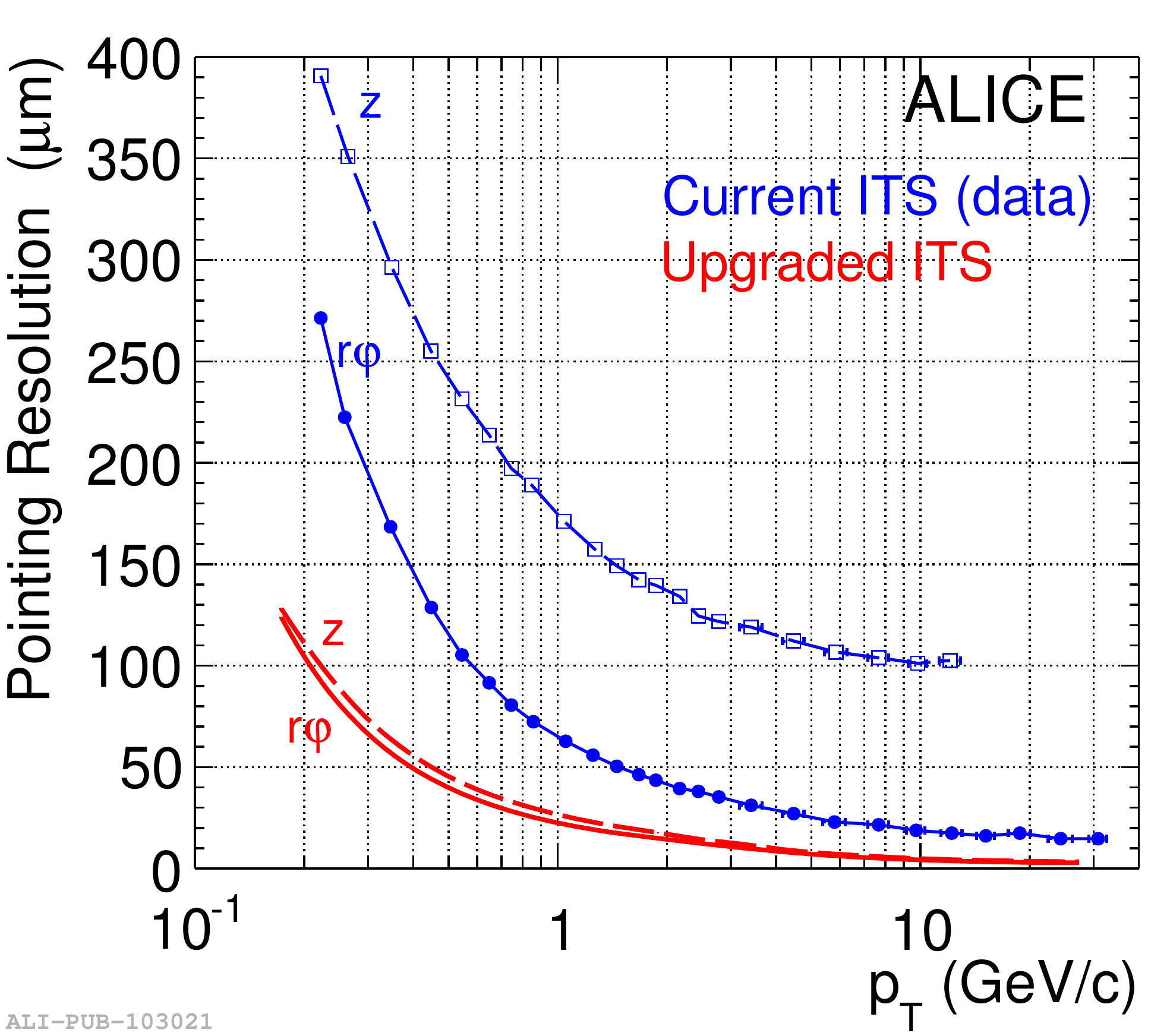}
\qquad
\includegraphics[width=.445\textwidth]{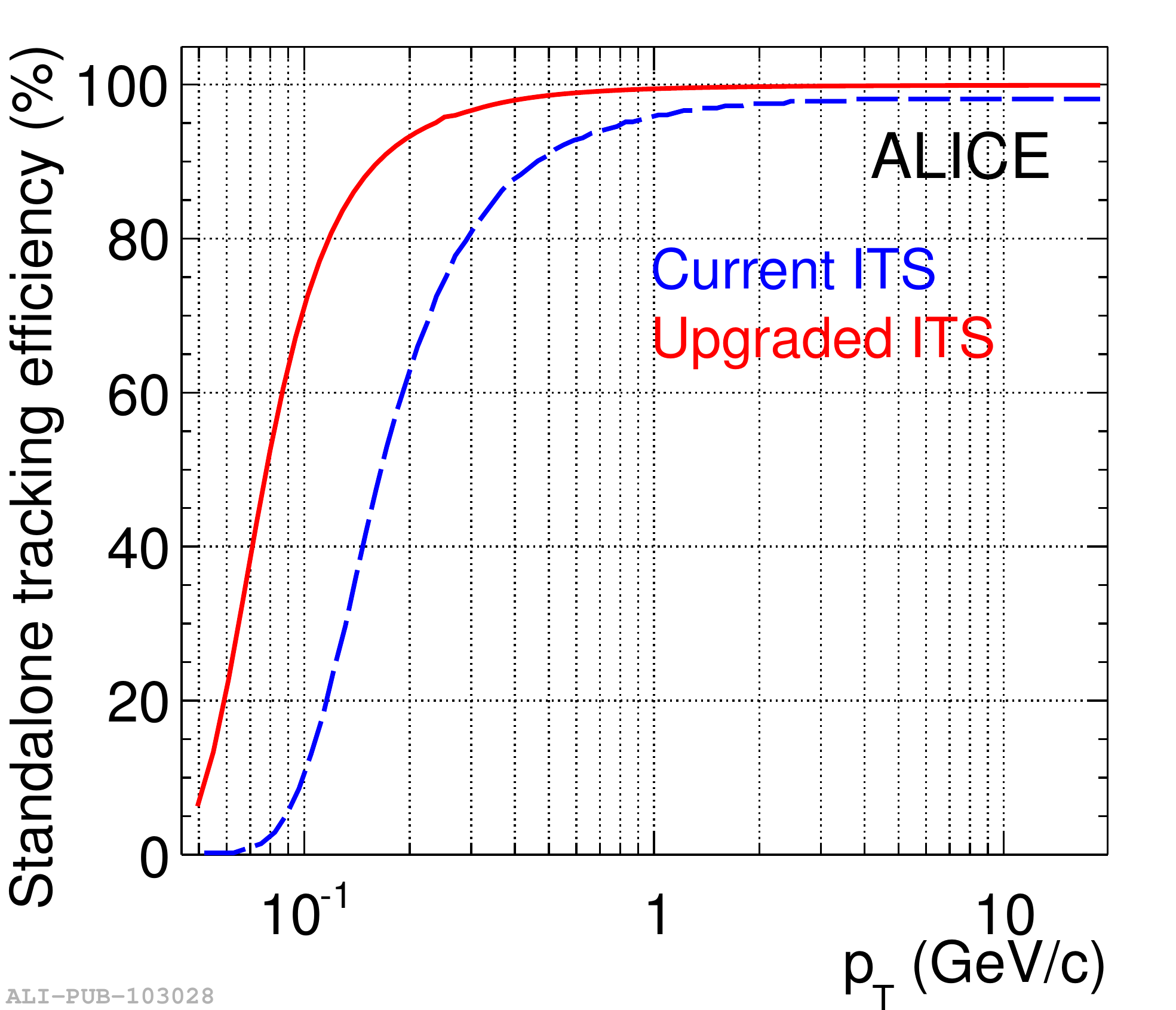}
\caption{\label{fig:efficiency}ITS2 simulated performance (red) compared with measured performance of the original ITS (blue). The track pointing resolution (left) and the tracking efficiency (right) are shown as a function of the tracked particle transverse momentum.}
\end{figure}

The ITS2 design characteristics 
will lead to a dramatic improvement of the ALICE detector tracking performance, especially for particles with transverse momentum smaller than 1~GeV/$c$.  
The material budget of the three innermost layers, as low as
0.35\% of a radiation length, combined with the ALPIDE spatial resolution of about 5~$\mu$m measured in beam tests and the 2.3~cm radius of the innermost layer, will drive the expected track impact parameter
resolution at $p_{\mathrm T}\sim$500~MeV/$c$ to about 40~$\mu$m, which improves the original tracker performance by a factor $\sim$3 in the transverse plane and $\sim$6 in the beam direction.
The new detector will also allow for an efficient track reconstruction
down to very low $p_{\mathrm T}$. The simulations show that the tracking efficiency at $p_{\mathrm T}\sim$0.1~GeV/$c$ will be improved by a factor $\sim$6 (see Fig.\ref{fig:efficiency}). 

The readout electronics will be able to record events at a typical rate of 50~kHz and a few 100~kHz for minimum bias Pb--Pb and pp collisions, respectively. 

\section{Detector production}
The ALPIDE chips have been tested and fully characterized after thinning and dicing with a custom-made automated probe-testing MAM\footnote{IBS Precision Engineering, http://www.ibspe.com/ (accessed 8 January 2020)} (Module Assembly Machine, also used for the assembly of the IB and OB modules), with a global yield of about 64\%.
The sensors are assembled in Hybrid Integrated Circuit (HIC) modules, which are in turn mounted on staves. The HIC and stave layouts have been designed in two different versions for the IB and the OB, respectively.

\begin{figure}[htbp]
\centering
\includegraphics[width=.5\textwidth]{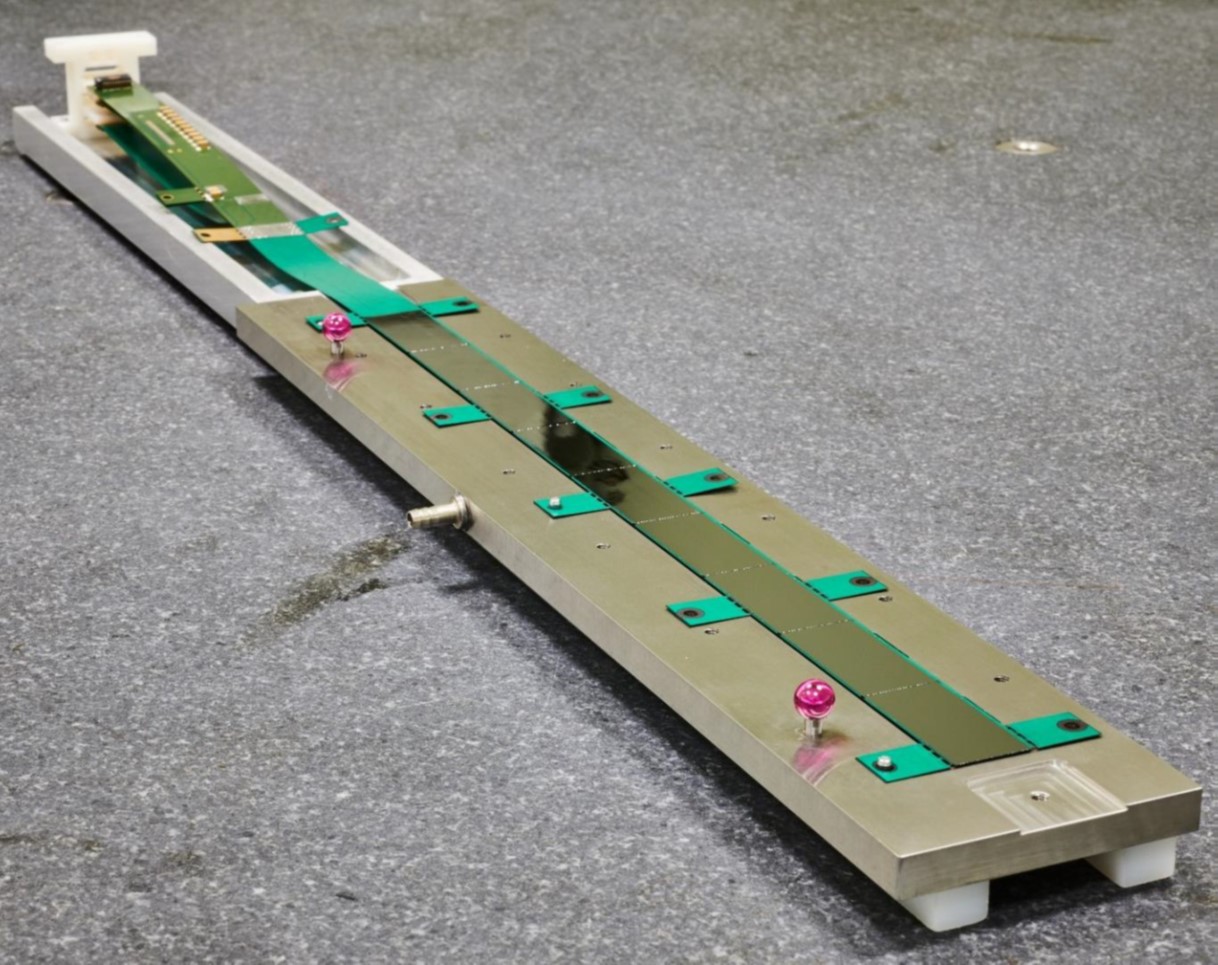}
\caption{\label{fig:IBHIC}The ITS2 Inner Barrel HIC, with the 9 ALPIDE chips facing upwards}
\end{figure}

The IB HIC, visible in Fig.\ref{fig:IBHIC}, consists of a single row of 9~$\times$~50~$\mu$m-thick ALPIDE chips aligned in the MAM and glued onto a fully validated aluminum-based Flexible Printed Circuit (FPC). The chip pads are wire-bonded to the FPC vias. Each chip will be then read out separately at 1.2~Gbps, in parallel with the others. The IB HIC is glued onto a carbon composite Cold Plate hosting two polyiamide cooling pipes, and on the Space Frame, a truss-like lightweight mechanical support structure. The global yields for the IB HIC and stave production resulted in 75\% and 97\%, respectively.
The 50~$\mu$m-thick ALPIDE chip testing campaign and the IB HIC and stave production have been carried out at CERN.

The OB HIC consists of 14~$\times$~100~$\mu$m-thick ALPIDE chips arranged in two rows: the first chip of each row acts as a master, transferring control signals and data to and from the other 6 chips on the same row. The chip pads corresponding to clock signal, control signal and data lines, are wire-bonded to a copper-based FPC. In the OB case, the power is delivered via 6 aluminum-Kapton cross-cables soldered directly to the FPC. The OB HIC production chain was shared among several institutes: the chips have been probe tested at the Pusan and Yonsei Institutes, in South Korea; the FPCs have been prepared and tested in Catania and Trieste, Italy; the HIC assembly was carried out in parallel in 5 different institutes (Bari in Italy,  Liverpool in UK, Pusan/Inha in South Korea, Strasbourg in France and Wuhan in China). The HIC production started in November 2017 and lasted 2 years, with a total number of 2679 assembled HICs, and a final detector-grade yield of 84\%, leaving 564 working spare OB HICs on top of the 1692 units composing the OB layers.

\begin{figure}[htbp]
\centering
\includegraphics[width=.8\textwidth]{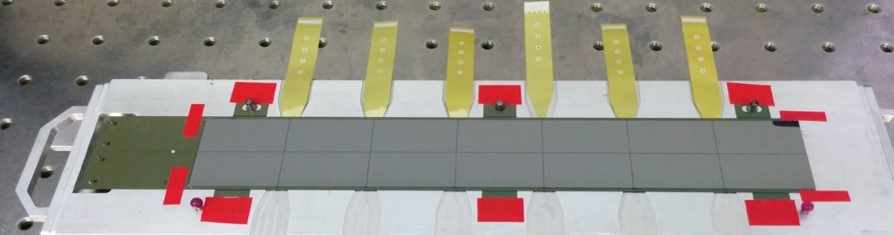}
\includegraphics[width=.8\textwidth]{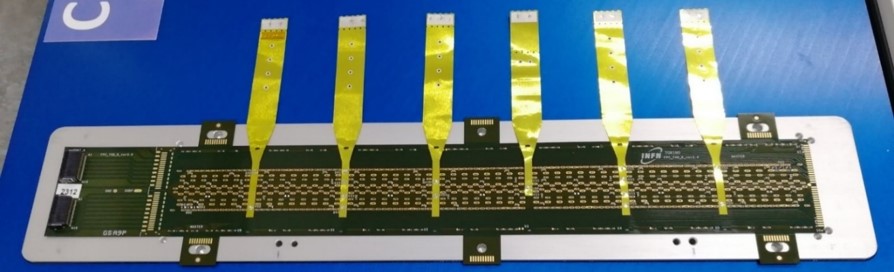}
\caption{\label{fig:OBHIC}The ITS2 Outer Barrel HIC, with the 14 ALPIDE chips facing upwards (top panel). View from the FPC side, with the six cross-cables soldered on it (bottom panel).}
\end{figure}

After full validation through a series of functionality tests, the OB HICs are shipped to the OB stave construction sites: Berkeley (USA) for the ML staves; Daresbury (UK), Frascati and Torino (Italy), Amsterdam (Netherlands) for the OL staves.
With the use of a Coordinate Measuring Machine (CMM)\footnote{Mitutoyo, https://www.mitutoyo.com/ (accessed 8 January 2020)}, a series of 4 (for the MLs) or 7 (for the OLs) HICs are aligned and glued onto a carbon composite cold plate, which embeds polyiamide cooling pipes. The HICs are electrically interconnected to each other by soldering conductive bridges on their short edges. An FPC extension is added to the first HIC to connect the full chain to the external readout and control systems. The unit so built is called Half-Stave (HS). 

\begin{figure}[htbp]
\centering
\includegraphics[width=.482\textwidth]{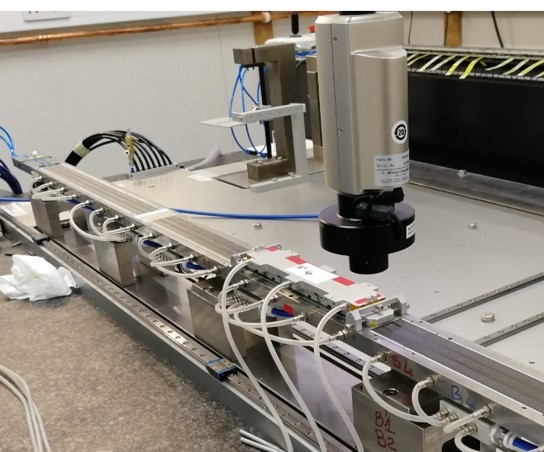}
\qquad
\includegraphics[width=.28\textwidth]{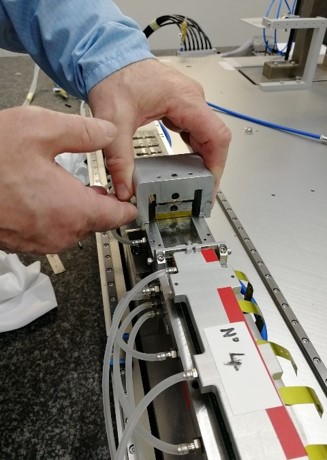}
\caption{\label{fig:OBStave}Half-stave assembly. A OB HIC is aligned on the cold plate with a CMM (left). The glue is then applied on the next cold plate segment with a spreading tool, in preparation for the installation of the adjacent HIC (right).}
\end{figure}

\begin{figure}[htbp]
\centering
\qquad
\includegraphics[width=.455\textwidth]{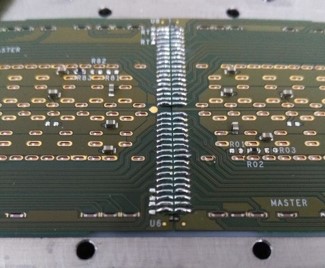}
\qquad
\includegraphics[width=.29\textwidth]{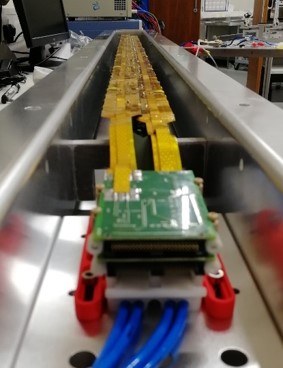}
\caption{\label{fig:PBUS}Left: detail of the interconnection bridge soldered to two adjacent HICs. Right: a full stave in the final compact layout, with the power bus installed and folded on top of the half-staves, boxed and ready to be shipped.}
\end{figure}

Two HSs are then aligned and glued with partial overlap onto a Space Frame with the corresponding length, in order to form a full OL stave. The cross-cables extending on the right and left sides of the corresponding HSs, are then soldered to an aluminum-Kapton power bus for power distribution. Full functionality tests have been performed throughout the construction procedure to monitor the quality of the produced devices and give immediate feedback in order to improve the assembly techniques. The material budget of each OB layer amounts to $\sim$1\% radiation length. The OB stave is designed to be read out at 400~Mpbs.
Selected operations from the OB stave assembly procedure are shown in Fig. \ref{fig:OBStave} and \ref{fig:PBUS}.

The OB stave production started in Summer 2018 and was completed in $\sim$1.5~years with a final detector-grade yield of $\sim$90\%, including the post-production rework of a few staves, initially presenting functional issues. A total number of 10 ML and 11 OL detector-grade spare staves have been produced, on top of the 54 ML and 90 OL staves needed for the detector assembly. All of them have been shipped to CERN and fully validated at the reception. 

The production of the ITS2 support structures and services~\cite{Piero}, including the readout electronics and the power system, has been completed and all the units have been delivered to CERN to be integrated with the detector.

\section{Commissioning at the surface}

\begin{figure}[htbp]
\centering
\includegraphics[width=.95\textwidth]{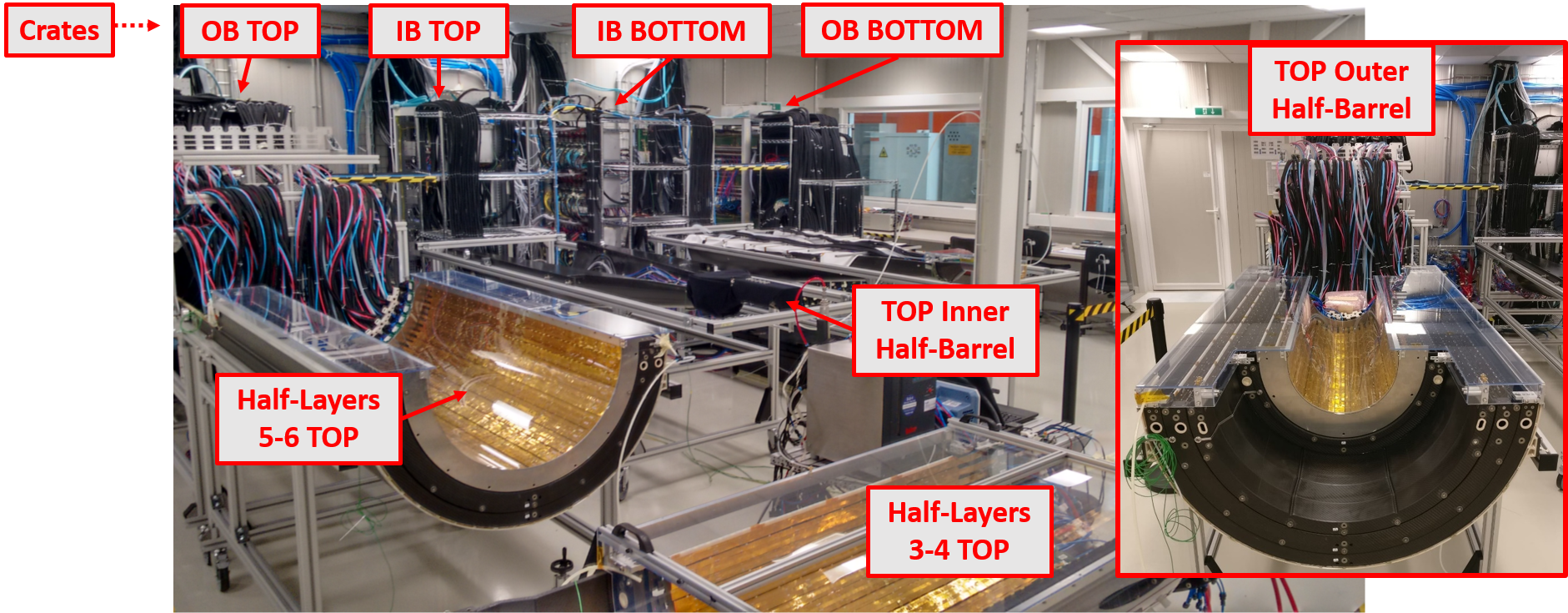}
\caption{\label{fig:CleanRoom}The four ITS2 half-barrels during the phase of commissioning at the surface in the CERN clean room. Each half-barrel is connected to the corresponding crate, hosting the power and readout system units. The right inset shows half-layer 3 and 4 installed on top of 5 and 6.}
\end{figure}

In a dedicated clean room at CERN, the staves have been mounted on the corresponding half-layer mechanical supports in order to build the 14 half-layers composing the ITS2 detector (see Fig.\ref{fig:CleanRoom}). The half-layers have been then progressively arranged together to form the two IB half-barrels and the two OB half-barrels. Starting from Summer 2019, the four units are being sequentially connected to the pre-tested cooling, readout, powering and control system elements, in order to validate the full detector chain and exercise the system.

\begin{figure}[htbp]
\centering
\includegraphics[width=.51\textwidth]{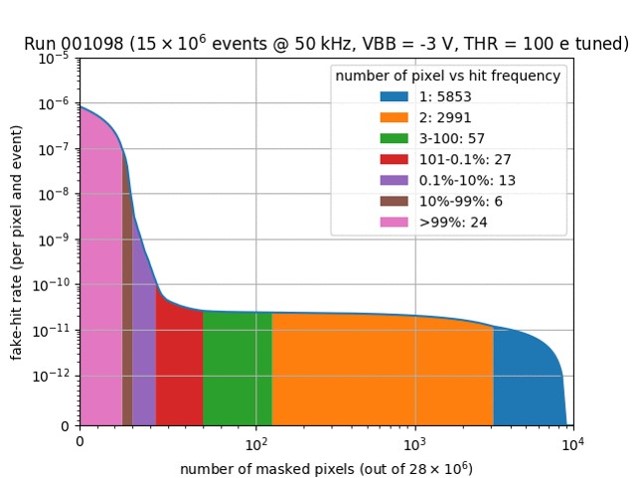}
\qquad
\includegraphics[width=.37\textwidth]{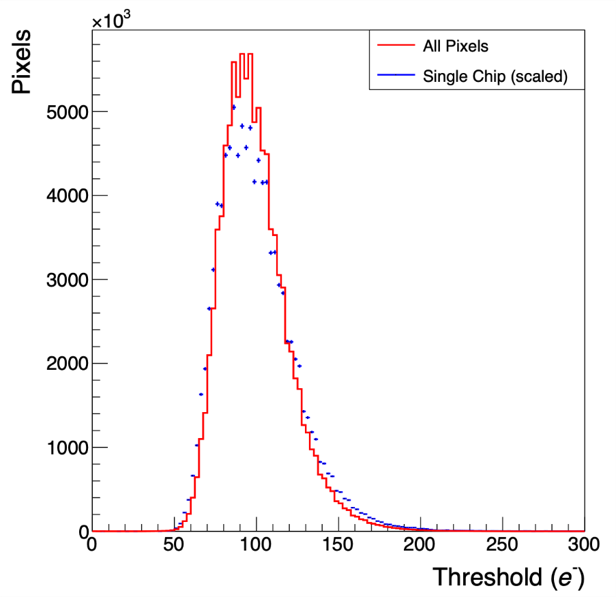}
\caption{\label{fig:FakeHit}Left panel: fake-hit rate as a function of the number of masked noisy pixels, measured on a spare IB half-layer. Right panel: pixel charge threshold distribution for a single chip (blue dots), scaled for comparison to the corresponding distribution for an entire OL stave (red line). The distributions overlap consistently, showing an RMS of $\sim$20~e$^-$.}
\end{figure}

In this configuration, the ITS2 half-barrels are undergoing a thorough phase of commissioning through noise, cosmic-ray, and radioactive source data-taking. From the noise data analysis, the detector appears to be extremely quiet: it shows a fake-hit rate of <~10$^{-10}$~ev$^{-1}$~pixel$^{-1}$ after masking only a fraction of noisy pixels smaller than 2$\times$10$^{-6}$ of the total read-out pixels (Fig.\ref{fig:FakeHit}).

The pixel-threshold spread and the noise have been measured on the $\sim$100~M pixels composing an OB stave: their distributions match the ones measured on a single chip, demonstrating that the calibration procedure and the detector response under the current conditions are well understood. Tuning the front-end parameters to equilibrate the charge thresholds allows for a uniform response across the detector, with a pixel charge threshold RMS as large as $\sim$20~e$^{-}$, reaching a very satisfying threshold stability over time. 

The commissioning at the surface will continue until Spring 2020, when the detector half-barrels, the electronics components and the cooling system will be transferred to the ALICE experimental site for final installation. The ITS2 will be then commissioned with the global systems for six months before the start of the Run3 operations.

\section{Conclusions}
After almost 10 years of fruitful operations and data-taking, the original ALICE Inner Tracking System has been decommissioned and extracted from the ALICE experimental site at the beginning of 2019~\cite{Elena}.

The ITS2, a new Inner Tracking System completely based on MAPS technology, has been developed with the aim of dramatically improving the ALICE tracking capabilities, especially at low momenta. The production of the ITS2 has been recently completed. The detector half-barrels are now undergoing a thorough commissioning at the surface, performing as expected in terms of fake-hit rate and stability. This phase will continue until Spring 2020, when the ITS2 will be installed in the ALICE detector and commissioned with the ALICE global systems before the LHC Run~3 data-taking, which is scheduled to start in 2021.

In the meantime, a further upgrade of the ITS Inner Barrel ("ITS3")~\cite{Magnus}, featuring minimal material budget and an even smaller radius of the innermost layer, has been proposed to be installed during the third LHC Long Shutdown (LS3), and the related R\&D activities have already started.

\end{document}